\documentclass[aps, pra, a4paper, twocolumn, 10pt, showpacs]{revtex4-1}
\usepackage{amsmath, amssymb, amsthm, bm, textcomp}
\usepackage[T1]{fontenc}
\usepackage[latin9]{inputenc}
\usepackage{graphicx,graphics}
\usepackage{geometry}
\usepackage{color}
\newcommand{\ket}[1]{\left|{#1}\right\rangle}
\newcommand{\bra}[1]{\left\langle{#1}\right|}

\newcommand{\be}{\begin{equation}}
\newcommand{\ee}{\end{equation}}
\newcommand{\eea}{\end{eqnarray}}
\newcommand{\bea}{\begin{eqnarray}}

\begin{document}

\title{Measurement-based quantum repeaters}

\author{M. Zwerger$^1$, W.\ D\"ur$^1$ and H. J. Briegel$^{1,2}$}

\affiliation{$^1$ Institut f\"ur Theoretische Physik, Universit\"at
  Innsbruck, Technikerstr. 25, A-6020 Innsbruck,  Austria.\\
  $^2$Institut f\"ur Quantenoptik und Quanteninformation der \"Osterreichischen Akademie der Wissenschaften, Innsbruck, Austria}
\date{\today}

\begin{abstract}
We introduce measurement-based quantum repeaters, where small-scale measurement-based quantum processors are used to perform entanglement purification and entanglement swapping in a long-range quantum communication protocol. In the scheme, pre-prepared entangled states stored at intermediate repeater stations are coupled with incoming photons by simple Bell-measurements, without the need of performing additional quantum gates or measurements. We show how to construct the required resource states, and how to minimize their size. We analyze the performance of the scheme under noise and imperfections, with focus on small-scale implementations involving entangled states of few qubits. We find measurement-based purification protocols with significantly improved noise thresholds. Furthermore we show that already resource states of small size suffice to significantly increase the maximal communication distance. We also discuss possible advantages of our scheme for different set-ups.
\end{abstract}

\pacs{03.67.Hk, 03.67.Lx, 03.67.-a}

\maketitle


\section{Introduction}
Quantum communication is one of the most advanced applications of quantum information processing, and a basic tool for quantum cryptography or distributed quantum computation. While the transmission of quantum information over distances of about a hundred kilometers has already been experimentally demonstrated by several groups, both in optical fibers \cite{Gi11} and free space \cite{Ur07}, long-range quantum communication over continental or intercontinental distances is still a challenging task. The concept of the quantum repeater introduced in \cite{Br98} provides in principle a scalable way for long-distance quantum communication, where the overhead in resources scales only polynomially with the distance. The quantum repeater combines entanglement swapping \cite{Be93,Zu93} and entanglement purification \cite{Be96,De96,Du07} ---i.e. the generation of high-fidelity entangled pairs from many copies of noisy pairs--- in such a way that high-fidelity entangled pairs are generated over large distances. These pairs can then be used for quantum communication or quantum cryptography, e.g. by teleportation \cite{Be93}. In the last decade, various  improvements and experimental proposals for quantum repeaters have been put forward \cite{Du01,Lo06,Ch06,Co07,Ji07,Be11}, and most of the required building blocks have already been experimentally demonstrated \cite{Pa03,Re06,Sh06,Yu08}. However, a full-scale quantum repeater still remains to be built, where the generation of entangled pairs, entanglement purification and entanglement swapping are combined with photon-matter interfaces in such a way that all components work with sufficiently high accuracy, i.e. with errors of a few percent.

Here we add a new element to such quantum repeater schemes, namely a measurement-based implementation of the required entanglement purification and entanglement swapping steps. While in conventional repeater proposals a sequence of coherent quantum gates and measurements performed on stored particles is required, in our scheme certain fixed entangled states are prepared and stored locally, and are then coupled with incoming photons directly via simple Bell measurements. No further coherent processing is required. One of the main advantages of such a scheme is that the preparation of the resource states can be done off-line, and even in a probabilistic way, without compromising the performance of the overall repeater scheme. In many set-ups, the preparation of certain fixed entangled states is significantly easier than achieving a coherent controlled manipulation of arbitrary quantum states by means of unitary gates.

We make use of results from measurement-based quantum computation \cite{Ra01,Br09}, where a fixed entangled state, the so-called cluster state \cite{Ra01b}, is used as a resource to perform universal quantum computation, i.e. an arbitrary sequence of unitary gates is realized by a sequence of single-qubit measurements performed on a sufficiently large 2D cluster state. As noted in \cite{Ra03}, a specific quantum algorithm or process can be obtained by using a specific, process-dependent resource state. In many cases, such a resource state will be a so-called graph state \cite{He04,He06}. The advantage of such a special-purpose measurement-based quantum processor is that it requires smaller states, i.e. fewer auxiliary qubits. Here we discuss such a construction for the operations involved in a repeater scheme. We show how to obtain the required resource states explicitly, and how to further reduce their size, where we in fact provide states of minimal size. As a by-product, we also obtain measurement-based entanglement purification protocols, where $n$ noisy entangled pairs are purified to yield (probabilistically) $m$ entangled pairs of higher fidelity. Remarkably this approach leads to a significantly larger error threshold of such purification schemes.

The paper is organized as follows. In Sec. \ref{SectionMQCprocessor} we describe two different methods of constructing special purpose quantum processors, which we apply in Sec. \ref{SectionMQCPurification} to obtain measurement-based entanglement purification and quantum repeater schemes with states of minimal size. We analyze the performance of such measurement-based quantum repeaters in the presence of noise and imperfections in Sec. \ref{SectionNoise}, and compare the achievable rates and distances when using quantum repeaters with small-scale quantum processors of limited size. We discuss possible advantages of using measurement-based elements in entanglement purification and repeater schemes for various set-ups in Sec. \ref{SectionSetups}, and summarize and conclude in Sec. \ref{SectionConlusion}.

\section{Background}
We first give a short introduction to some important concepts and schemes which we will need later. In the context of measurement-based quantum computation the notion of graph states significantly simplifies the analysis. In addition we describe the quantum repeater and its building blocks, entanglement purification and swapping.

\subsection{Graph states}

We start with reviewing graph states and their properties \cite{He04,He06}. A graph $G$ is a pair
\bea
G=(V,E)
\eea
where $V\subset \mathbb{N}$ and $E\subset [V]^2$. $V$ and $E$ are finite sets, their elements are called vertices and edges, respectively. Here we restrict to simple graphs which do neither contain loops (edges connecting vertices with itself) nor multiple edges between the same vertices. The neighborhood $N_a$ of a vertex $a\in V$ is the set of vertices $b$ with $\{a,b\} \in E$.

Here we associate with any graph $G=(V,E)$ a graph state $\ket{G}$ in a Hilbert space $\mathcal{H}_V=(\mathbb{C}^2)^{\otimes V}$ in the following way: for any $a\in V$ one defines
\bea
K_G^{(a)}=X^{(a)} \prod_{b\in N_a} Z^{(b)}.
\eea
$X, Y, Z$ denote the Pauli matrices and the superscript refers to the particle/Hilbert space on which the operator acts. The graph state $\ket{G}$ is then defined as the common eigenvector with unit eigenvalue for all $a\in V$,
\bea
K_G^{(a)}\ket{G}=\ket{G}.
\eea
Alternatively $\ket{G}$ can be written as
\bea
\ket{G}=\prod_{\{a,b\}\in E} CZ^{(a,b)}\ket{+}^{\otimes V}
\eea
where $CZ^{(a,b)}$ is the controlled $Z$ operation on the vertices $a$ and $b$, i.e. $CZ^{(a,b)}=diag(1,1,1,-1)$.

Two graph states $\ket{G}$ and $\ket{G'}$ with associated graphs $G=(V,E)$ and $G'=(V,E')$, respectively, are called local unitary (LU) equivalent, if there exists a local unitary $U$ such that \cite{He04}
\bea
\ket{G}=U\ket{G'}.
\eea
An important tool in the context of graph states is local complementation, which is very useful if one considers the effect of Pauli measurements on a graph state. To describe this concept, we first introduce a few definitions.

From $V'\subset V$ and a graph $G=(V,E)$ one can obtain a new graph denoted by $G-V'$ by deleting all vertices $V'$ and all edges which are incident with an element of $V'$. Likewise $G-E'$ refers to the graph where one deletes all edges $e\in E'$ with $E'\subset E$ and for general $F\subset [V]^2$ we write $G+F=(V,E\cup F)$. With the symmetric difference of $E$ and $F$,
\bea
E\triangle F=(E\cup F)-(E\cap F)
\eea
one defines $G\triangle F=(V,E\triangle F)$.

Furthermore, let
\bea
E(A,B)=\{\{a,b\} \in E: a \in A, b\in B, a\neq b\}
\eea
denote the set of edges between two sets $A, B\subset V$ of vertices.
Local complementation is now defined as the transformation $\tau_a: G \to \tau_a(G)$, where the edge set $E'$ of $\tau_a(G)$ is given by $E'=E\triangle E(N_a,N_a)$. The graph states $\ket{G}$ and $\ket{\tau_a{G}}$ are LU equivalent: $\ket{\tau_a{G}}=U_a(G)\ket{G}$ with
\bea
U_a(G)=\left(-iX^{(a)}\right)^{1/2} \prod_{b\in N_a} \left(iZ^{(b)} \right)^{1/2}.
\eea
With local complementation at hand one can derive simple rules for the effect of a Pauli measurement on a graph state, which will become important below in the context of measurement-based quantum computation. Let
\bea
P_{z,\pm}^{(a)}=\frac{1\pm Z^{(a)}}{2}
\eea
denote the projector onto the eigenvector $\ket{z,\pm}$ with eigenvalue $\pm 1$ and similar for $X$ and $Y$.
Then \cite{He04,He06}
\bea
\label{Paulirules}
P_{z,\pm}^{(a)}\ket{G}&=& \ket{z,\pm}^{(a)} \otimes U_{z,\pm}^{(a)} \ket{G-a} \nonumber\\
P_{y,\pm}^{(a)}\ket{G}&=& \ket{y,\pm}^{(a)} \otimes U_{y,\pm}^{(a)} \ket{\tau_a(G)-a} \\
P_{x,\pm}^{(a)}\ket{G}&=& \ket{x,\pm}^{(a)} \otimes U_{x,\pm}^{(a)} \ket{\tau_{b_0}\left( \tau_a \circ \tau_{b_0}(G)-a\right)} \nonumber
\eea
where the $LU$ operations $U_{x,\pm}^{(a)}$ etc. can be found in the appendix and $b_0$ is an arbitrary vertex in $N_a$.

\subsection{Quantum repeaters}
The ability to efficiently establish entangled Bell pairs over large distances is a key element for quantum networks and distributed quantum computation. One way to achieve this is provided by the quantum repeater \cite{Br98}.
The concept of a quantum repeater allows the creation of entangled pairs with only polynomial overhead in the distance. To achieve this, one combines entanglement purification and swapping together with a quantum memory. Here we use a so-called recurrence protocol for entanglement purification. It enables the creation of an entangled pair with higher fidelity from several pairs of lower fidelity via local operations. Each party performs single and two qubit gates as well as measurements such that one pair remains. The measurement outcomes determine whether the purification was successful. There exist protocols for various numbers of input and output pairs. The output pairs of a protocol can also be used as inputs again, until the desired fidelity is reached.

Entanglement swapping enables to create an entangled pair between parties A and C given Bell pairs between parties A and B, and between B and C, via a Bell measurement on the two qubits at location B. It is conceptually close to quantum state teleportation. Unless the Bell pairs and the measurement are both perfect, the swapping will lower the fidelity of the final pair. The extension to several pairs is straightforward.

In the quantum repeater scheme one divides a channel of length $l$ into $N$ segments (for simplicity it is assumed that $N=L^n$ for some integer $n$) and establishes entangled pairs over each of them. $M$ pairs are used to obtain a pair of higher fidelity using entanglement purification, where the precise value of $M$ depends on the protocol, the values of the initial and final fidelity and the noise of the involved operations. Then $L$ pairs are connected via entanglement swapping so that a Bell pair over a larger distance is established. As long as the fidelity after the swapping remains above the minimal required fidelity for the purification protocol, one can iterate this procedure until all segments are connected. Notice that connection with subsequent purification to the initial fidelity defines one repeater level, where one faces exactly the same situation as before, however with pairs of larger and larger distance.
The total number $R$ of elementary pairs is given by
\bea
R=N^{\log_L M+1}
\eea
which shows the polynomial dependence on $N$ and therefore the distance $l\propto N$. For a more detailed description we refer the reader to \cite{Br98}.

\subsection{Clifford group}

The Clifford group plays a crucial role in entanglement purification protocols and many quantum error correcting codes, as many of them solely contain Clifford gates. It is generated by the $CZ$, Hadamard and $U_z(\pi /2)=\exp(-i \pi/4 Z)$ gates. Changing the order of a Clifford operation $C$ and a Pauli operator leaves the Clifford operation unchanged, i.e. $C\sigma=\sigma'C$, where $\sigma$ and $\sigma'$ are possibly different tensor products of Pauli operators.


\section{Small-scale measurement-based processors}\label{SectionMQCprocessor}
In this section, we discuss two methods how to construct quantum states that can be used to perform a specific quantum process, i.e. a unitary circuit or a combination of a unitary circuit and measurements. The first method is based on the Jamiolkowski isomorphism \cite{Ja72,Ci01}, while the second method uses the universality of the 2D cluster state together with simplifications of the state for specific measurements \cite{Ra03}.

\subsection{Jamiolkowski isomorphism}

The Jamiolkowski isomorphism relates a completely positive map (CPM) and a state \cite{Ja72}. Here we illustrate it for a map on two qudits, but the generalization is straightforward. Given a CPM ${\cal{E}}$ acting on systems $A_1$ and $B_1$ with $d$-dimensional Hilbert spaces $\mathcal{H}_{A_1}$ and $\mathcal{H}_{B_1}$, respectively, the associated state is
 \bea
E_{A_1A_2,B_1B_2}={\cal{E}}(P_{A_1A_2}\otimes P_{B_1B_2})
\eea
with
\bea
P_{A_1A_2}=\ket{\phi}_{A_1A_2}\bra{\phi}
\eea
where
\bea
\ket{\phi}_{A_1A_2}=\frac{1}{\sqrt{d}}\sum_{i=1}^{d}\ket{i}_{A_1}\otimes \ket{i}_{A_2}
\eea
is a maximally entangled state.
In the following we restrict to qubits so that $d=2$ and $\ket{\phi}_{A_1A_2}$ is simply the Bell state $\ket{\phi^+}=(|00\rangle + |11\rangle)/\sqrt{2}$. With $E_{A_1A_2,B_1B_2}$ at hand one can implement the map ${\cal E}$ on any state of systems $A_3$ and $B_3$ by performing a Bell measurement $\{|\phi_k\rangle\}$ on $A_2A_3$ and $B_2B_3$ \cite{Ci01}, where $|\phi_k\rangle = \mathbb{I} \otimes \sigma_k |\phi^+\rangle$ with Pauli operators $\sigma_0=\mathbb{I}, \sigma_1= X, \sigma_2 = Y, \sigma_3 = Z$. However if one projects on a Bell state different from $\ket{\phi_0}$ one has to deal with a Pauli operator that is efficiently implemented before the actual map $\cal{E}$. For measurement outcomes corresponding to $|\phi_i\rangle$ in $A$ and $|\phi_j\rangle$ in $B$, one obtains $\rho_{out}={\cal{E}}\left(\left(\sigma_i\otimes \sigma_j\right) \rho_{in} \left(\sigma_i\otimes \sigma_j\right)\right)$. In general this means that the implementation has failed, unless one can commute the Pauli operator $(\sigma_i\otimes \sigma_j)$ and the map $\cal{E}$ without changing the map ${\cal E}$ (but possibly the Pauli operator), and hence obtain the desired operation up to local Pauli corrections. For unitary $\cal{E}$ this is equivalent to $\cal{E}$ being in the Clifford group, which is the case for all the protocols we consider. That is, any unitary in the Clifford group can be implemented deterministically. In order to relate $E_{A_1A_2,B_1B_2}$ with a graph state we compute the rank indices which uniquely identify the graph states considered in this work \cite{He04,Ca09}. The local unitaries (LU), which are not specified by this procedure can be determined via a numerical optimization.

\subsection{Measurement-based quantum computation}

The idea of measurement-based quantum computation \cite{Ra01,Br09} is to initially prepare a large entangled state, a so called resource state. A quantum circuit is then translated to a (single-qubit) measurement pattern which implements the quantum operation. There are several resource states which allow for universal quantum computation, the most prominent is the 2D cluster state \cite{Ra01b}. It is a graph state with the topology of a two dimensional square lattice. The read-in of the initial state, on which the computation shall be performed, can be done via joint Bell measurements with the cluster state.

In order to obtain a resource state for a particular map, one can start with a sufficiently large 2D cluster state and the measurement pattern for the desired operation. The effect of any Pauli measurement included in the measurement pattern on a graph state is to reduce the number of (connected) vertices by at least one and to change the edge structure and the local unitary operations. The resulting state is obviously again a graph state as can be seen from the transformation rules above, however it is no longer universal. In measurement-based quantum computation all Pauli measurements can be done beforehand and therefore the size of the graph state is reduced (see \cite{Ra03}). In this context it is important to note that any Clifford gate can be implemented with Pauli measurements only. Consequently any circuit with $n$ inputs and $m$ outputs containing only Clifford operations and Pauli measurements can be realized deterministically on a $n+m$ qubit graph state. The graph state can be explicitly determined by sequentially applying the rules (\ref{Paulirules}).

%

\section{Measurement-based entanglement purification and quantum repeaters}\label{SectionMQCPurification}
In this section we show how to apply the construction of Sec. \ref{SectionMQCprocessor} to obtain states that can be used to perform entanglement purification \cite{Be96,De96,Du07}, or the combination of entanglement purification and entanglement swapping.


We use the Oxford protocol \cite{De96} for entanglement purification because of its fast convergence. It operates on two noisy Bell pairs as input. Rotations around the x-axis by an angle of $\tfrac{\pi}{2}$ on  one side and $-\tfrac{\pi}{2}$ on the other are followed by a bilateral $CNOT$ gate and measurements of the target qubits in the computational basis. The map is thus given by ${\cal E}\rho= O \rho O^\dagger$ with
\begin{widetext}
\bea
{O}=\left(P_{z}^{(A_t)}\otimes P_{z}^{(B_t)}\right) \left( CNOT^{(A_{c\rightarrow  t})}\otimes CNOT^{(B_{c\rightarrow  t})}\right) \left(e^{-\tfrac{i\pi}{4}{Z^{(A)}}^{\otimes 2}}\otimes e^{+\tfrac{i\pi}{4}{Z^{(B)}}^{\otimes 2}}\right)
\eea
\end{widetext}
where the superscripts $A, B$ refer to the two parties and $c, t$ to the control and target qubit, respectively.

The purification is successful when both measurement outcomes coincide. Since all required unitary operations are in the Clifford group they can be implemented with  Pauli measurements only so that the final graph state  only has qubits for the input and the output. There are no single-qubit measurements any more, the whole computation is finally executed by the Bell measurements at the read-in. {

In the case of one entanglement purification step there are two inputs and one output (on each side) and therefore the graph state will have three qubits. It can easily be determined up to LU operations since there is only one fully connected three-qubit graph state, the GHZ state:
\bea
\ket{GHZ_3}=\frac{1}{\sqrt{2}}\left(\ket{000}+\ket{111}\right).
\eea
In general one starts with the whole circuit for the desired map, which may contain several rounds of purification and swapping depending on the purpose, and determines the associated Jamiolkowski state. With this state one identifies the graph state and the LU operations as described above.

An additional key ingredient for a quantum repeater is entanglement swapping. It is simply done by a projection on a Bell state, which can be decomposed into a Hadamard and $CNOT$ gate, which is in the Clifford group since it can be constructed from two Hadamard and one $CZ$ gate,  followed by Pauli measurements. Thus in a measurement-based approach it can be implemented with Pauli measurements only, which consequently leads to compact resource states.
The resource state which performs two sides of the purification protocol and entanglement swapping in one step (cf. figure (\ref{FigEPP2})) has four input and no output qubits and therefore is a four qubit graph state.

The graph states for one and two rounds of entanglement purification as well as for entanglement purification and integrated swapping are shown in figures (\ref{FigEPP2}) and (\ref{FigEPP4}). The superscripts $+$ and $-$ in the resource states refer to the different local transformations in the purification protocol at the left and right side, respectively. The LU operations, which are determined via optimizing the overlap with the desired resource state from the Jamiolkowski isomporphism can be found in the appendix. Note that they are not unique, but the different solutions only lead to global phases. In the protocol for the derivation of the graph states it is assumed that the target qubits are projected on the $\ket{0}$ state. However this does not mean that the purification is always successful, as will be explained below.


\begin{figure}[htb]
\centering
\includegraphics[scale=0.5]{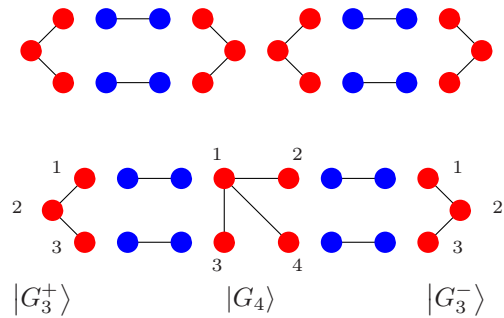}
\scriptsize
\put(-165,30){1}
\put(-180,15){2}
\put(-165,0){3}
\put(-105,35){1}
\put(-75,35){2}
\put(-105,-7){3}
\put(-75,-7){4}
\put(-15,30){1}
\put(0,15){2}
\put(-15,0){3}
\normalsize
\put(-180,-20){$\ket{G_3^{+}}$}
\put(-100,-20){$\ket{G_4}$}
\put(-25,-20){$\ket{G_3^{-}}$}
\caption{Resource states (red) for one purification step without (top) and with (bottom) integrated entanglement swapping acting on input Bell pairs (blue). When successful, one long distance purified output pair is generated from four short distance Bell pairs shared between $A-C$ and $C-B$ respectively. Notice that the resource states $|G_3^+\rangle$ and $|G_3^-\rangle$ at the left and right end of the Bell pairs are different due to the asymmetry of the Oxford purification protocol.}
\label{FigEPP2}
\end{figure}

\begin{figure}[htb]
\centering
\includegraphics[scale=0.5]{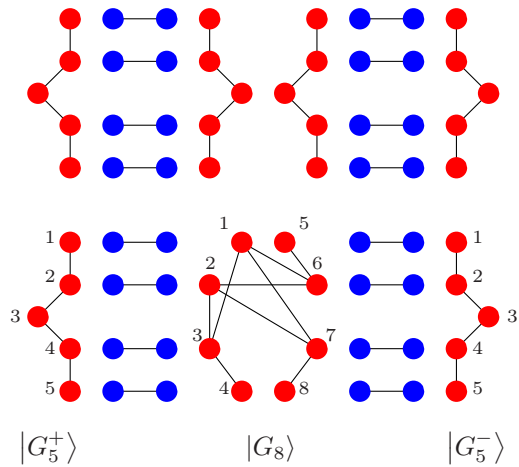}
\scriptsize
\put(-170,60){1}
\put(-170,45){2}
\put(-183,30){3}
\put(-170,18){4}
\put(-170,3){5}
\put(-105,65){1}
\put(-110,50){2}
\put(-75,65){5}
\put(-70,50){6}
\put(-115,23){3}
\put(-65,23){7}
\put(-105,3){4}
\put(-75,3){8}
\put(-10,60){1}
\put(-10,45){2}
\put(3,30){3}
\put(-10,18){4}
\put(-10,3){5}
\normalsize
\put(-180,-20){$\ket{G_5^{+}}$}
\put(-95,-20){$\ket{G_8}$}
\put(-20,-20){$\ket{G_5^{-}}$}
\caption{Resource states (red) for two purification steps without (top) and with (bottom) integrated entanglement swapping, acting on input Bell pairs (blue). When successful, one long distance purified output pair is generated. Notice that the resource states $|G_5^+\rangle$ and $|G_5^-\rangle$ at the left and right end of the Bell pairs are different due to the asymmetry of the Oxford purification protocol. }
\label{FigEPP4}
\end{figure}


An important issue is the read in, done by a joint Bell measurement on the qubits of the noisy Bell pairs and the graph state. Whenever one does not project on the Bell state $\ket{\phi^{+}}=\tfrac{1}{\sqrt{2}}\left(\ket{00}+\ket{11}\right)$ one has to deal with an undesired Pauli byproduct operator. However, since the protocol involves only unitary gates in the Clifford group and Pauli measurements the byproduct operator can be commuted through, leading to a Pauli byproduct operator for the final pair. In addition the measurement outcomes determine which projection on the target qubits has in fact been implemented. Recall that only cases with same measurement outcomes at $A$ and $B$ correspond to a successful purification, which is hence determined by the outcome of the Bell measurements. This is due to the fact that the Pauli operator from the Bell measurement also has to be commuted through the projection $P_0$ from the purification protocol and thereby possibly changes it to $P_1$. From table \ref{Tabmeasurements} one can determine the basis correction for the final resulting pair as well as the actually implemented projections  for all possible projections at the read in: any possible byproduct operator can be generated from the four byproduct operators shown in the table. For concatenated purification the rules simply have to be applied repeatedly. We now illustrate this with an example (cf. figure (\ref{FigMeasurements})):

\begin{table}[htb]
\caption{Logical table from which effective measurement outcome and correction operation can be determined. The first operator of the input acts on the control qubit and the second on the target qubit. $P_{0/1}$ denotes projection on the $\ket{0}/\ket{1}$ state. When concatenated, the Pauli operators are simply multiplied, whereas for the projections the following rule applies: $P_i\circ P_j=P_{i\oplus j}$, where $\oplus$ denotes addition modulo 2. It is illustrated in more detail in an example.}
\vspace{0.5cm}
\centering
\begin{tabular}{c c c}
\hline\hline
input & effective projection & output \\
\hline
$\mathbb{I}\otimes X$ & $P_1$ & $\mathbb{I}$ \\
$\mathbb{I}\otimes Z$ & $P_1$ & $Z$ \\
$X \otimes \mathbb{I}$ & $P_1$ & $X$ \\
$Z \otimes \mathbb{I}$ & $P_1$ & $ZX$ \\
\hline
\end{tabular}
\label{Tabmeasurements}
\end{table}


\begin{figure}[htb]
\centering
\includegraphics[scale=0.5]{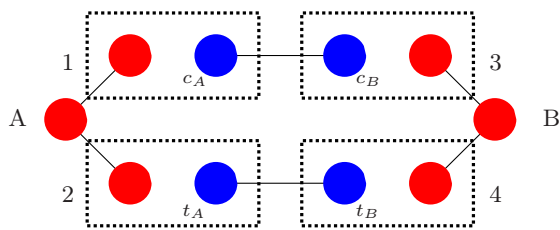}
\put(-170,60){1}
\put(-170,10){2}
\put(-10,60){3}
\put(-10,10){4}
\put(-190,39){A}
\put(10,39){B}
\scriptsize
\put(-125,55){$c_A$}
\put(-125,5){$t_A$}
\put(-60,55){$c_B$}
\put(-60,5){$t_B$}
\caption{Example for the application of table \ref{Tabmeasurements}.  One purification step is considered. The dashed boxes represent Bell measurements. For the details see main text.}
\label{FigMeasurements}
\end{figure}

Assume that the qubits in box 1 are projected on $\ket{\psi^+}=\left(\mathbb{I}\otimes X\right)\ket{\phi^+}$, the ones in box 2 on $\ket{\phi^-}=\left(\mathbb{I}\otimes Z\right)\ket{\phi^+}$ and the two others on $\ket{\phi^+}$. In this case on the right hand side (party $B$) the desired protocol with the projection $P_0$ and without byproduct operator is implemented. However on the left hand side (party $A$) one has two Pauli operators on the input: $X$ on the control ($c_A$) and $Z$ on the target ($t_A$) qubit so that the rules from table \ref{Tabmeasurements} have to be applied repeatedly. The $X$ on the control gives a $X$ on the output and the $Z$ on the target gives a  $Z$ on the output so that the final byproduct operator is $XZ$ which determines the local basis of the resulting Bell pair. For the projection one finds $P_1\circ P_1=P_{1\oplus 1}=P_0$. Since the right hand side is also projected on $\ket{0}$, this corresponds to a successful purification step.

The same can be done for the $4 \to 1$ entanglement purification protocol shown in figure (\ref{FigEPP4}). There, two subsequent purification rounds on 4 initial Bell pairs are combined into a single operation, leading eventually to one purified output pair. Again, the results of the Bell measurements determine whether the overall procedure (i.e. all three purification processes) was successful, and at the same time determine the local correction operation. One simply takes the local correction operators from the first two purifications as the input for the third purification step (cf. table \ref{Tabmeasurements}).

There also exist other purification protocols with only Clifford gates and Pauli measurements that map $n$ pairs to $m$ pairs, e.g those based on quantum error correcting codes introduced in \cite{As05}. These can be implemented on a $n+m$ qubit graph state as described above and also require only Bell measurements and no additional single qubit measurements. The same holds for schemes which combine quantum error correction and entanglement purification \cite{Be11} as long as only Clifford gates and Pauli measurements are involved.

\section{Influence of noise and imperfections}\label{SectionNoise}

\subsection{Entanglement purification protocol}

In any realistic implementation the quantum operations will be imperfect. Here we investigate the influence of noise on the measurement-based entanglement purification protocol. We consider the influence of imperfect resource states. To this aim, we use a simple but rather general noise model, where we assume that each individual particle of the resource state is subjected to local white noise (LWN). This captures both the imperfect generation of the resource state, and possible memory errors during storage. Notice that within this model, the fidelity of larger resource states decreases with system size, thereby capturing the fact that the generation of larger resource states is increasingly difficult and or involves a larger number of operations. We also assume that the Bell measurements are perfect, however an imperfect Bell measurement can simply be included in the LWN of the resource states, leading to a higher value of LWN.

Given an $n$-qubit density matrix $\rho$, the noisy density matrix reads
\bea
\rho_{LWN}=\left(\prod_{j=1}^{n}{\cal{D}}_j\right)\rho
\eea
with
\bea
{\cal{D}}_j\rho=p\rho + \frac{1-p}{2}\mathbb{I}_j\otimes \operatorname{tr}_j\rho
\eea
where $p\in[0,1]$ quantifies the level of noise.


We now use this error model to determine the noise threshold for measurement-based entanglement purification using noisy resource states. The LWN threshold, i.e. the maximal LWN value such that the fidelity of noisy Bell pairs still increases if one purifies them with the special resource states as described above, is determined numerically. We find a threshold of $(1-p_1)=3.5\%$ for one purification step ($\ket{G_3}$) and $(1-p_2)=7.1\%$ for the protocol which concatenates two purification steps ($\ket{G_5}$). This corresponds to a fidelity, i.e. overlap with the desired state, of 92.3\% and 76.1\%, respectively. Notice that our results differ from the usual gate based approach, where one has noise thresholds for the single and two qubit gates, which do not change when the protocol is concatenated. Intuitively this can be understood in the following way: two purification steps can be done either with three 3-qubit states or with one 5-qubit state on each side. In the latter case the noise is distributed on only five qubits instead of nine, which leads to the improved error threshold and has no analogue in the gate based approach. Thus the possibility to combine several purification steps and perform them with a resource state of smaller size allows one to obtain a significantly improved error threshold together with a more compact implementation. However this new protocol will have a lower success probability as explained below.


\subsection{Quantum repeater protocol}
\label{probabilityreduction}
Imperfect operations also affect the required overall resources in the quantum repeater scheme. Explicitly the resources $R$ are given by
\bea
R=N^{\log_L M+1}
\eea
for any quantum repeater independent of the specific implementation. Here $N$ is proportional to the distance, $L$ is the number of pairs which are connected at each repeater level (here we use $L=2$) and $M$ is the number of pairs one needs to obtain one pair of higher fidelity. The value of $M$ is determined by the explicit purification protocol, the noise of the involved quantum operations and their success probability. The scheme proposed here requires many Bell measurements, so that the overall performance will strongly depend on their success. Let $p_{Bell}$ denote the success probability for a Bell measurement. Then $M\propto p_{Bell}^{-2m}$ for a $m \rightarrow 1$ protocol, which clearly shows the importance of a high probability $p_{Bell}$.

In the case of simultaneous purification and swapping the total success probability is lower, since one has to require that the purification for both of the pairs which are connected is successful in the same step. Consequently one will need more attempts and thus more resources (photon pairs). The same holds for the concatenation of two purification rounds where all three purification steps have to be successful simultaneously. So there will be a trade-off between minimizing the distributed resources and the locally stored qubits.

\section{Comparison of rates and achievable distances}\label{SectionRates}

In order to determine the achievable rates we use the model introduced in \cite{Co05} to calculate the fidelity of a photonic Bell pair transmitted through optical fibres as a function of distance. The fidelity $F$ for a binary pair, i.e. a mixture of $\rho_{\ket{\phi^+}}$ and $\rho_{\ket{\phi^-}}$, is given by
\bea
F=\frac{1+V}{2}
\eea
where $V$ is given by
\bea
V=\frac{V_{opt}^2 \left( t\eta \left(1-D\right) \right) ^2}{\left( \left(t\eta+\left(1-t\eta \right)2D\right) \left(1-D\right) \right) ^2}
\eea
where $V_{opt}$ is the optical visibility, $\eta$ is the detector efficiency and $D$ its dark count rate. The probability of a photon to pass a given distance of $d$ km without being absorbed, $t$, is given by
\bea
t=10^{-\alpha d/10}
\eea
so that there are losses of $\alpha$ dB/km. Here we have chosen $V_{opt}=0.99$, $\eta=0.3$ and $D=10^{-4}$ as in \cite{Co05}, whereas for the damping we assume $\alpha=0.16$. Notice that considering more optimistic parameters leads to a significant reduction of required resources.
We then take a binary state with this fidelity as our input for the quantum repeater protocol. Any other kind of noise such as noisy Bell measurements is included in the LWN per particle of the resource states. This model allows to obtain an estimate of the achievable rates and fidelities independent of the particular physical setup, but should be replaced by a more accurate one once one considers a specific physical implementation.

The resulting fidelities for various numbers of repeater levels are displayed in figures (\ref{FigDistance1}) and (\ref{FigDistance2}). Here the length of the elementary segments is varied so that one obtains a continuous plot. It can be seen that for low noise (order of one percent of local white noise per particle) one purification step per level is enough and that with two purification steps per level comparatively high noise (several percent) can be tolerated. Notice that one can achieve higher fidelities by performing more entanglement purification steps per repeater level, at the price of an enlarged overhead. In addition we find that six or at most seven repeater levels suffice to reach the intercontinental scale.


\begin{table}[htb]
\caption{Achievable fidelity and required resources (total number of elementary pairs) for different number of repeater levels when considering one purification step per repeater level with integrated swapping. Local white noise $(1-p)$ of 1\% per particle for $\ket{G_3}$ and $\ket{G_4}$.}
\vspace{0.5cm}
\centering
\begin{tabular}{c c c c}
\hline\hline
\# levels & distance [km] & fidelity & overhead \\
\hline
$3$ & $1000$ & $95.40\%$ & $1.42\cdot10^5$ \\
$4$ & $1000$ & $95.40\%$ & $3.48\cdot10^3$ \\
$5$ & $5000$ & $92.48\%$ & $2.13\cdot10^7$ \\
$6$ & $5000$ & $94.76\%$ & $8.34\cdot10^4$ \\
$6$ & $10000$ & $91.88\%$ & $6.90\cdot10^7$ \\
$7$ &$ 10000$ & $94.50\%$ & $2.35\cdot10^5$ \\
$8$ &$ 20000$ & $94.26\%$ & $6.75\cdot10^5$ \\
\hline
\end{tabular}
\label{Tabdistance1}
\end{table}

\begin{table}[htb]
\caption{Achievable fidelity and required resources (total number of elementary pairs) for different number of repeater levels when considering two purification steps per repeater level with integrated swapping. Local white noise $(1-p)$ of 4\% per particle for $\ket{G_5}$ and $\ket{G_8}$.}
\vspace{0.5cm}
\centering
\begin{tabular}{c c c c}
\hline\hline
\# levels & distance [km] & fidelity & overhead \\
\hline
$3$ & $1000$ & $91.81\%$ & $1.10\cdot10^7$ \\
$4$ & $1000$ & $91.63\%$ & $4.36\cdot10^6$ \\
$5$ & $5000$ & $90.98\%$ & $4.55\cdot10^{10}$ \\
$6$ & $5000$ & $91.25\%$ & $9.99\cdot10^{8}$ \\
$6$ & $10000$ & $90.95\%$ & $7.41\cdot10^{11}$ \\
$7$ & $10000$ & $91.14\%$ & $1.57\cdot10^{10}$ \\
$8$ &$ 20000$ & $91.07\%$ & $2.30\cdot10^{11}$ \\
\hline
\end{tabular}
\label{Tabdistance2}
\end{table}

The overheads, i.e. the number of pairs that have to be sent through the elementary segments in order to obtain one pair on the full distance, can be found in tables \ref{Tabdistance1} and \ref{Tabdistance2}. It can be seen that one can lower the overheads by introducing more repeater levels, i.e. by making the elementary segments smaller. Note that the overheads can be reduced if the purification and swapping are not done in the same step, i.e. by using only the resource states $\ket{G_3}$ and $\ket{G_5}$ (a further reduction for the case of two purification steps is possible if one uses three $\ket{G_3}$ states instead of $\ket{G_5}$). Other possibilities to significantly increase the rate are parallelization and multiplexing \cite{Co07}. In certain cases one accepts a rate of order of one photon per minute \cite{Co05}. Assuming a standard photon source pumped at 10 GHz this then corresponds to an overhead of $6\cdot 10^{11}$.

%
%

\begin{figure}[htb]
\centering
\includegraphics[scale=0.5]{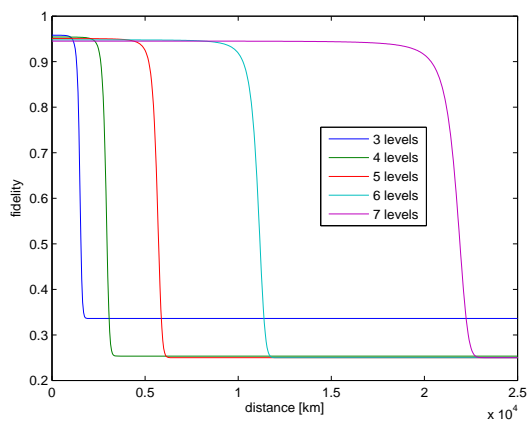}
\caption{Achievable fidelity as a function of distance when considering one purification step per repeater level with integrated swapping (cf. figure (\ref{FigEPP2})). Local white noise $(1-p)$ of 1\% per particle for $\ket{G_3}$ and $\ket{G_4}$.}
\label{FigDistance1}
\end{figure}

\begin{figure}[htb]
\centering
\includegraphics[scale=0.5]{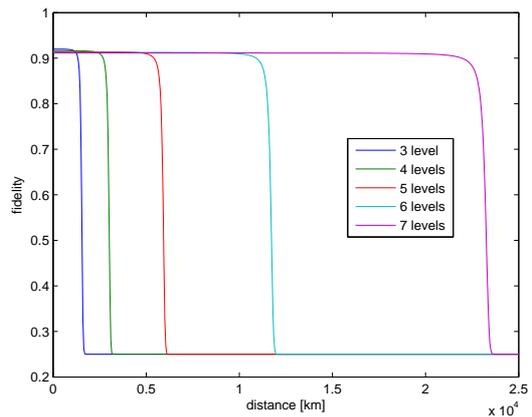}
\caption{Achievable fidelity as a function of distance when considering two purification steps per repeater level with integrated swapping (cf. figure (\ref{FigEPP4})). Local white noise $(1-p)$ of 4\% per particle for $\ket{G_5}$ and $\ket{G_8}$.}
\label{FigDistance2}
\end{figure}

\section{Measurement-based repeater schemes for different experimental set-ups}\label{SectionSetups}
In this section we discuss different variants of measurement-based repeater schemes and possible advantages of a measurement-based scheme as compared to a gate-based approach.

\subsection{Variants of measurement-based quantum repeaters}
We start by discussing different variants of measurement-based repeater schemes. As pointed out in the context of entanglement purification schemes and entanglement swapping (see Sec. \ref{SectionMQCPurification}), one can reduce the size of the resource states such that only input and output particles are kept. This remains true also if one combines two subsequent purification rounds, or an entanglement purification step followed by entanglement swapping. However, the reduction of size yields to smaller success probabilities as we will discuss in the following (see also Sec. \ref{probabilityreduction}). We compare three different possibilities.

(V1): If one considers two subsequent purification rounds operating on four entangled input states, the first purification step involves two three--qubit states at each party (each resource state consisting of two input particles and one output particle), and another three-qubit resource state for the second purification step. In addition, the output particles of the first purification round need to be coupled to the input particles of the resource state for the second round by means of Bell measurements. This involves a total of nine particles at each site, but has the advantage that if one of the purification steps is not successful, the remaining particles (entangled pairs) are still unharmed and can be further used, e.g. be combined with other pairs where first-round purification was successful. In addition, several independent resource states are used, which consist of fewer particles as compared to resource states used in (V2) and (V3) (see below).

(V2): One can combine the two subsequent purification rounds (three purification processes) into a single one, where now four input particles are mapped onto one output particle at each site, i.e. the procedure involves a five-qubit resource state. Notice that one has a significant reduction of size of the resource state and there is no need to couple output particles of the first round to other resource states by means of Bell measurements, however all three purification attempts need to be simultaneously successful. This leads to an overall smaller success probability as compared to (V1).

(V3): One can combine to some extent the advantages of schemes (V1) and (V2). On the one hand, one can reduce the number of particles as compared to scheme (V1), but keep the advantage of re-using remaining particles/pairs if some purification step is not successful. On the other hand only single qubit measurements are involved (except the initial Bell measurements that couple particles of the Bell pairs to input particles of the resource state), in contrast to additional Bell measurements required to connect output and input particles of different rounds in (V1). This is achieved as follows: we consider the step-by-step measurement-based realization of subsequent purification rounds. For scheme (V3) all measurements (except that on input and output particles) are performed on the initial (enlarged) resource state. This is possible because all operations belong to the Clifford group. However, one could also not perform some of these measurements. This is exactly what we do here, namely we keep intermediate particles that correspond to output states of the first purification round (these particles coincide with the input particles for the second purification round). If the first purification round for one of the pairs turns out to be not successful, one can simply erase the corresponding output particles and decouple them from the remaining system. Notice that this is always possible, as states are local unitary equivalent to graph states where single particles can be erased and decoupled from the remaining system by $Z$-measurements \cite{He04}. In this way, the remaining particles which may e.g. correspond to a purified Bell pair can be further used. Notice that it is also possible to consider larger states that involve not only four initial Bell pairs, but six or more in such a way that more than two output particles of pairs resulting from the first purification round can be combined to participate in the second purification round (see figures (\ref{FigV1}), (\ref{FigV2}), (\ref{FigV3}) for a schematic illustration). This has the advantage that even if one of the first purification steps is not successful, a second purification step with two output particles of a successful first purification step can be performed by means of single-qubit measurements only, and no coupling between resource state of different rounds by means of Bell measurements is required.

Notice that similar variants are possible also for other involved operations, including e.g. entanglement purification plus subsequent entanglement swapping, or several subsequent rounds of entanglement purification (e.g. an eight to one protocol corresponding to three rounds).
If we look at the overall repeater scheme, in particular at higher levels, there is also the need to combine output pairs from lower levels and process them further using additional resource states. Again, this can be done with different variants (V1),(V2),(V3). While (V1) requires independent resource states of small size but a larger number of total particles together with additional coupling by means of Bell-measurements, one could use enlarged resource states similar as in (V2) and (V3).
Notice that  using (V2) at all levels, i.e. states of minimum size which combine the purification and swap operations, leads to a small success probability. In particular this implies that all purification steps need to succeed simultaneously. This destroys the polynomial scaling of the repeater scheme for the overall resources and implies that schemes (V1) or (V3) need to be used, e.g. at higher repeater levels.

\begin{figure}
\centering
\includegraphics[scale=0.4]{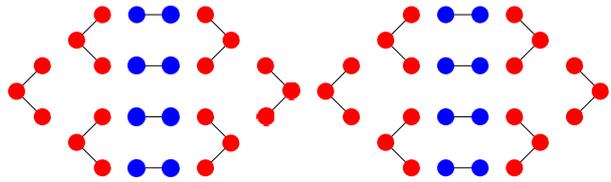}
\caption{Illustration of version V1. The schematic setup for two purification steps and one entanglement swapping is shown. Particles of Bell pairs (blue) are coupled via Bell measurements to neighboring particles of the resource state (red), where output particles of the first purification round are further coupled to the resource states that perform the second round of purification. Additional Bell measurement for entanglement swapping in the middle is required.}
\label{FigV1}
\end{figure}

\begin{figure}
\centering
\includegraphics[scale=0.5]{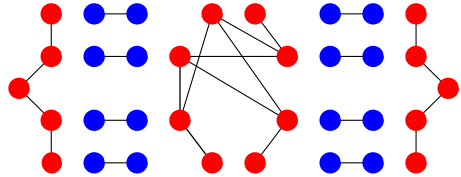}
\caption{Illustration of version V2. Same setup as in figure (\ref{FigV1})  but resource states of minimal size which combine all purification and swapping steps.}
\label{FigV2}
\end{figure}
%

\begin{figure}[htb]
\centering
\includegraphics[scale=0.5]{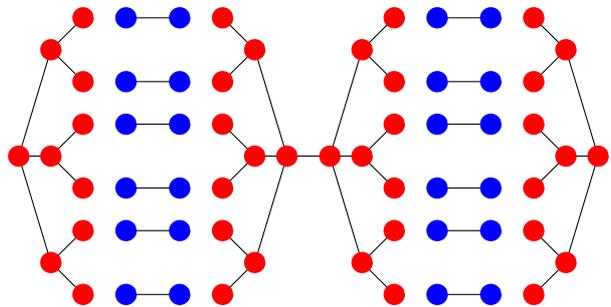}
\caption{Illustration of version V3. Same setup as in figure (\ref{FigV1}).  Here one can allow for one unsuccessful purification step in the first round. Notice that the figure only provides a schematic illustration and may not correspond to the actual graph states that are required to implement the desired operation.}
\label{FigV3}
\end{figure}

\subsection{Advantages of measurement-based schemes for different set-ups}
Here we discuss possible advantages of measurement-based repeater schemes for different set-ups.

First we note that the measurement-based repeater scheme solely relies on two ingredients: (i) the generation and local storage of certain entangled states, all of which are equivalent --up to a local basis change-- to graph states; (ii) Bell measurements between stored particles and incoming photons or (depending on the used scheme) two stored particles. In particular, no coherent two-qubit operations or other interactions are required. Regarding (i), we point out that the generation of the entangled states can be done in a probabilistic fashion and beforehand, i.e. before the transmission of photons through channels takes place. The states only need to be available once transmitted photons arrive. This does not only allow for probabilistic generation schemes, as e.g. discussed or implemented in the context of photons [parametric down conversion, cluster state generation of photons], but offers also the possibility to use entanglement purification schemes to increase the fidelity of the pre-prepared states. It is however crucial that the states can be stored locally \cite{Ha07}. In principle, all entangled states could be generated by means of photons (via parametric down conversion, beam splitters and passive optical elements), where then the state of each of the photons is transferred to another quantum information carrier, e.g. an ion, an atom, a quantum dot or an atomic ensemble, where the quantum state can then reliably be stored. To this aim an interface between photons and matter is required (see \cite{Ha10,Lv09} and also e.g. \cite{Interface,QMemory} for recent developments in this direction), which allows to transfer the state of the photon to the storage system. One possibility is that the atomic system is placed into a high-finesse cavity, and the photonic state is then mapped to the cavity mode and subsequently onto a single atomic system, but other schemes such as a storage in an atomic ensemble or a BEC \cite{Le11} are also possible.

Notice that whenever one uses scheme (V3), it is in principle sufficient to have only one storage qubit per cavity, as no interaction between storage qubits is required \cite{footnote}. In addition, it is not necessary to transfer the state of the storage qubit to a photonic state, i.e. a one-way interface is sufficient. This significantly simplifies the requirements of such measurement-based schemes as compared to gate-based approaches.

In certain set-ups, it might however be easier to work with two storage qubits, one for storing the particle of the resource state and one for mapping the incoming photonic state onto, where a Bell measurement is then performed on the two storage qubits after the successful coupling process. This makes it easier to implement coupling schemes that allow one to detect a successful transfer from photonic to atomic degrees of freedom, or to determine whether a photon actually arrived. Notice that it is important to only perform measurements on the stored particle -which might carry a state resulting from many previous purification and entanglement swapping processes- when the photon has arrived and the state is successfully mapped to the storage unit. Otherwise, the advantage of a quantum memory would be lost.

What seems particularly attractive is that only (probabilistic) photonic entanglement sources suffice, together with interfaces and storage units that only involve a single qubit per storage unit. We remark however that an "all photonic" realization without quantum memory leads to extremely large overheads, simply due to the fact that all photons have to arrive simultaneously at all repeater nodes. However, only very few additional purification steps suffice to achieve continental or even intercontinental distances (see Sec. \ref{SectionRates}).

\section{Summary and Conclusions}\label{SectionConlusion}

In this work, we presented a measurement-based implementation of quantum repeaters for long-distance quantum communication and distributed quantum computation. The proposed scheme requires only two- and possibly single-qubit measurements and can integrate several rounds of entanglement purification and swapping into a single step. This integration leads to smaller resource states, i.e. states with less ancilla qubits, and in addition to significantly higher error thresholds. Furthermore we found that for reasonable levels of noise one or at most two entanglement purification steps per repeater level, and six levels in total, suffice to reach the intercontinental scale.

We also discussed some possible experimental implementations and their advantages. Here it is appealing that the generation of the resource states can be probabilistic without affecting the scheme's performance. Another interesting aspect is that quantum memories which can store only single qubits can be sufficient to implement this repeater scheme without the need of entangling gates between them.



\section*{Acknowledgements}
This work was supported by the Austrian Science Fund (FWF): SFB F40-FoQus F4012-N16, P20748-N16, and the European Union (NAMEQUAM).


\appendix

\section{ LU operations}

The LU operations for the various resource states shown in figures (\ref{FigEPP2}),(\ref{FigEPP4}) are given in table \ref{tableLU}.\\
The LU that appear in the transformation rules for graph states under Pauli measurements are given by

\bea
U_{z,+}^{(a)}&=&\mathbb{I}\\
U_{z,-}^{(a)}&=&\prod_{b\in N_{a}} Z^{(b)}\\
U_{y,+}^{(a)}&=&\prod_{b\in N_{a}}\left(-iZ^{(b)}\right)^{1/2}\\
U_{y,-}^{(a)}&=&\prod_{b\in N_{a}}\left(iZ^{(b)}\right)^{1/2}\\
U_{x,+}^{(a)}&=&\left(iY^{(b_0)}\right)^{1/2} \prod_{b\in N_a -N_{b_0}-\{b_0\}} Z^{(b)}\\
U_{x,-}^{(a)}&=&\left(-iY^{(b_0)}\right)^{1/2} \prod_{b\in N_{b_0}-N_a-\{a\}} Z^{(b)}
\eea

\begin{widetext}

\begin{table}[b]
\label{tableLU}
\caption{LU operations for resource states.}
\vspace{0.5cm}
\centering
\begin{tabular}{c c c c c c c}
\hline\hline
 qubit \# & $\ket{G_{3}^{+}}$ & $\ket{G_{3}^{-}}$ & $\ket{G_4}$ & $\ket{G_{5}^{+}}$ & $\ket{G_{5}^{-}}$ & $\ket{G_8} $ \\
 \hline
$1$ & $(iZ)^{\tfrac{1}{2}}$ & $(iZ)^{\tfrac{1}{2}}$ & $(iX)^{\tfrac{1}{2}}$ & $(iZ)^{\tfrac{1}{2}}$ & $(iZ)^{\tfrac{1}{2}}$ & $(-iZ)^{\tfrac{1}{2}}$ \\
$2$ & $\mathbb{I}$ & $X$ & $(-iZ)^{\tfrac{1}{2}}(iX)^{\tfrac{1}{2}}$ & $Z(iX)^{\tfrac{1}{2}}$ & $Z(iX)^{\tfrac{1}{2}}$ & $(iZ)^{\tfrac{1}{2}}(iX)^{\tfrac{1}{2}}$ \\
$3$ & $X(-iZ)^{\tfrac{1}{2}}$ & $X(-iZ)^{\tfrac{1}{2}}$ & $(iZ)^{\tfrac{1}{2}}(iX)^{\tfrac{1}{2}}$ & $\mathbb{I}$ & $Z$ & $(iX)^{\tfrac{1}{2}}$ \\
$4$ & & & $(iZ)^{\tfrac{1}{2}}(-iX)^{\tfrac{1}{2}}$ & $(iX)^{\tfrac{1}{2}}$ & $(-iX)^{\tfrac{1}{2}}$ & $X(iZ)^{\tfrac{1}{2}}$ \\
$5$ & & & & $X(iZ)^{\tfrac{1}{2}}$ & $X(-iZ)^{\tfrac{1}{2}}$ & $(iZ)^{\tfrac{1}{2}}$ \\
$6$ & & & & & & $Z(iX)^{\tfrac{1}{2}}$ \\
$7$ & & & & & & $(-iX)^{\tfrac{1}{2}}$ \\
$8$ & & & & & & $X(-iZ)^{\tfrac{1}{2}}$ \\
\hline
\end{tabular}
\end{table}

\end{widetext}


\begin{thebibliography}{99}
\bibitem{footnote}
The same is true for scheme (V2) within a single repeater level, but coupling between different repeater levels involves additional Bell measurements.
\end{thebibliography}
\end{document}